# Inversion of Magnetotelluric Data using Bayesian Neural Networks

Dhruv Poddar [*1], Rohan Sharma[1], Divakar Vashisth[2]

[1]*Department of Applied Geophysics, Indian Institute of Technology, Dhanbad,* [2]*Department of Energy Science and Engineering, Stanford University*

email: * dhruvpoddar2016@gmail.com

## Summary

Magnetotelluric (MT) inversion is a key technique in geophysics for imaging deep subsurface resistivity structures. However, the inherent ill-posedness and non-uniqueness of inverse problems make them challenging to solve. While supervised deep learning approaches have shown promise in this domain, their predictions are typically deterministic and fail to capture the associated uncertainty, an essential factor for decision-making. To address this limitation, we explore the application of Bayesian Neural Networks (BNNs) for MT inversion with uncertainty quantification. Specifically, we train a Bayesian Convolutional Neural Network (BCNN) on a synthetically generated MT dataset. The BCNN effectively recovers resistivity profiles from apparent resistivity data, with the predicted means closely matching the ground truth across both the training and test sets, while also providing uncertainty estimates quantified within ±3 standard deviations from the mean. These results underscore the potential of BNNs to enhance deep learning-based geophysical inversion frameworks by incorporating principled uncertainty quantification.

## Introduction

Magnetotelluric (MT) inversion aids in imaging the Earth's subsurface electrical resistivity structure using naturally occurring electromagnetic fields (Simpson & Bahr, 2005). By analyzing variations in the Earth's electric and magnetic fields over a wide range of frequencies, MT surveys provide valuable information about conductivity changes at varying depths, spanning near-surface geological formations to deeper crustal and mantle features. However, recovering resistivity profiles from MT data constitutes a challenging inverse problem due to its inherently ill-posed, non-linear, and non-unique nature (Constable et al., 1987), necessitating sophisticated computational frameworks for stable and interpretable solutions. Traditionally, MT inversion has been approached through various algorithms, such as Occam's inversion (Constable et al., 1987), Gauss-Newton methods (Siripunvaraporn & Egbert, 2000), and particle swarm optimization (Shaw & Srivastava, 2007).

In recent years, deep learning–based MT inversion (Singh et al., 2019; Liao et al., 2022; Xie et al., 2023; Yu et al., 2025) has emerged as a powerful alternative, offering the ability to model the complex, non-linear mapping between apparent resistivity data and subsurface resistivity profiles without requiring explicit physical formulations. A key advantage of these methods lies in their efficiency: once trained, the model can produce near-instantaneous inversions for new datasets, eliminating the need to re-run traditional solvers. However, a major limitation of conventional deep learning approaches is their deterministic nature—they yield point estimates without capturing the uncertainty inherent in geophysical inverse problems. Moreover, deep neural networks are known to produce overconfident predictions, particularly when trained on limited data (Goodfellow et al., 2016). Given the ill-posedness of geophysical inversion and its susceptibility to data and model uncertainties, incorporating uncertainty quantification becomes critical. Uncertainty estimates not only enhance the interpretability of inversion results but also support risk-aware decision-making in exploration workflows (Zou et al., 2021). To address these challenges, we propose a Bayesian Neural Network (BNN) framework for supervised MT inversion that enables both accurate prediction and principled quantification of uncertainty.

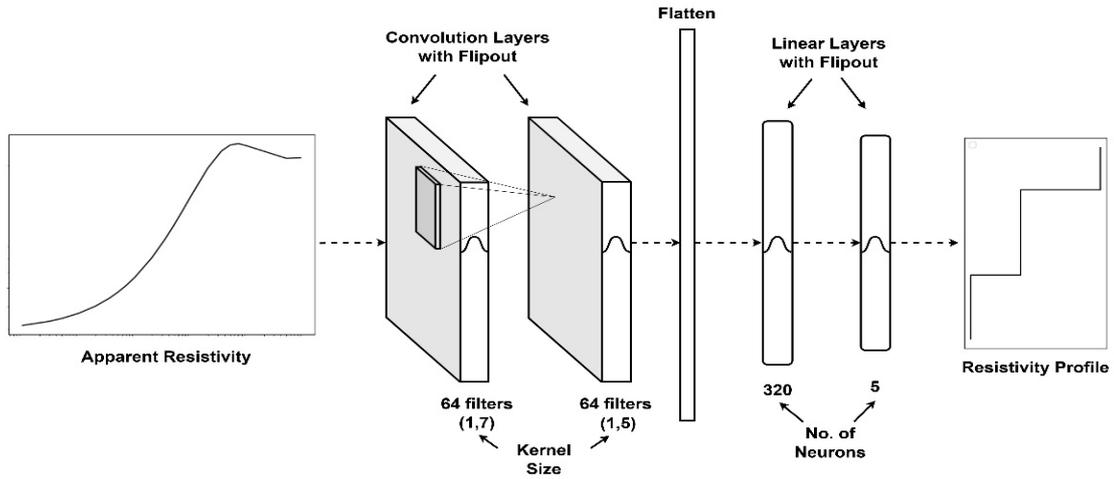

*Figure 1* *Architecture of the Bayesian Convolutional Neural Network (BCNN) model used to estimate resistivity profiles and associated uncertainties from MT data.*

## Methodology

The proposed methodology employs a BNN to learn the underlying relationship between apparent resistivity measurements and their corresponding resistivity profiles. Unlike traditional neural networks, which use fixed weights, BNNs model these weights as probability distributions (Blundell et al., 2015), thereby enabling the quantification of uncertainty in their predictions. Among the various methods available for learning these weight distributions, this study adopts a Variational Inference (VI)–based BNN approach (Li et al., 2024). In VI, instead of directly computing the true posterior distribution over the weights, which is often intractable—a simpler family of distributions, referred to as the variational family, is defined (Zhao et al. 2024). The goal is to identify the member of this family that most closely approximates the true posterior. This approximation is achieved by minimizing a divergence measure between the true and variational distributions, with the Kullback-Leibler (KL) divergence (Kullback & Leibler, 1951) being commonly used for this purpose.

*Table 1* *Parameter search space used for dataset generation, along with RMSE (between the predicted mean and ground truth) and associated uncertainty (±3 standard deviations) for each subsurface layer on the training and test sets.*

|  | Search Space | | Results on Training Set | | Results on Test Set | |
|---|---|---|---|---|---|---|
|  | Minimum | Maximum | RMSE | Uncertainty | RMSE | Uncertainty |
| $\rho_1$ (Ωm) | 10,000 | 50,000 | 942.287 | 2683.299 | 978.508 | 2580.06 |
| $\rho_2$ (Ωm) | 100 | 25,000 | 1166.525 | 1861.806 | 1341.464 | 1858.821 |
| $\rho_3$ (Ωm) | 1 | 5000 | 104.364 | 346.242 | 117.612 | 347.334 |
| $t_1$ (m) | 5000 | 25,000 | 1145.337 | 1497.954 | 1254.395 | 1465.158 |
| $t_2$ (m) | 10,000 | 25,000 | 1221.094 | 1331.916 | 1405.237 | 1356.186 |

During training, the BNN samples a set of weights from the learned variational distributions, effectively generating a deterministic neural network instance. The input apparent resistivity is passed through this network to predict a resistivity profile, which is then compared to the ground truth using the mean squared error (MSE). Additionally, the KL divergence between the variational and prior distributions over the weights is computed, scaled by a factor of 0.01, and added to the MSE to form the total loss function. This combined objective guides the training process by balancing prediction accuracy and posterior regularization. For inference, weights are sampled multiple times (1000 in our case) from the learned distributions to effectively generate an ensemble of neural networks. Each network, when given the same input apparent resistivity data, produces a different resistivity profile prediction. The mean of these ensemble of outputs is taken as the final predicted profile, while the spread—quantified using ±3 standard deviations—provides an estimate of uncertainty.

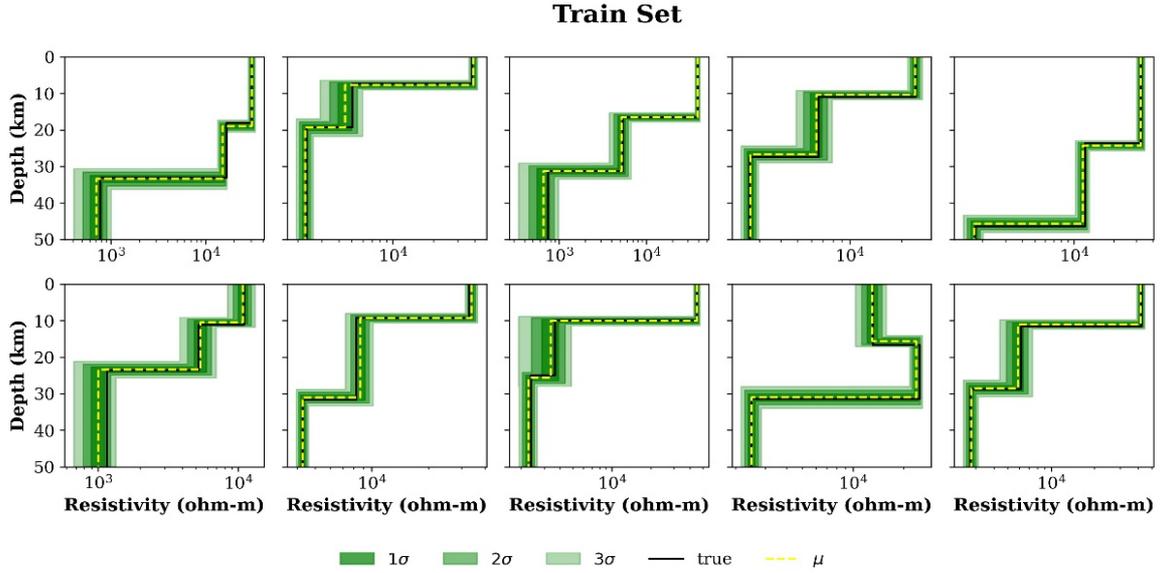

**Figure 2** Estimated resistivity profiles from the BCNN model (Figure 1) for selected training set examples. True profiles are shown in black, mean predictions in yellow, and green shaded regions represent uncertainty bounds corresponding to ±3 standard deviations around the mean.

In this study, we implement a Bayesian Convolutional Neural Network (BCNN) architecture (Figure 1). The BCNN employs Bayesian convolutional and linear layers with the Flipout estimator (Wen et al., 2018). The proposed architecture comprises two convolutional layers followed by a linear layer with Sigmoid and ReLU activations, respectively, and concludes with a final linear output layer without any activation. The number of nodes in the output layer corresponds to the number of model parameters to be predicted—in this case, five. To train the model, a synthetic dataset comprising 30,000 three-layer resistivity profiles and their corresponding apparent resistivity curves was generated using the parameter search space defined in Sarkar et al. (2023), as summarized in Table 1. Eighty-five percent of the dataset was allocated for training and validation, with the remainder reserved for testing.

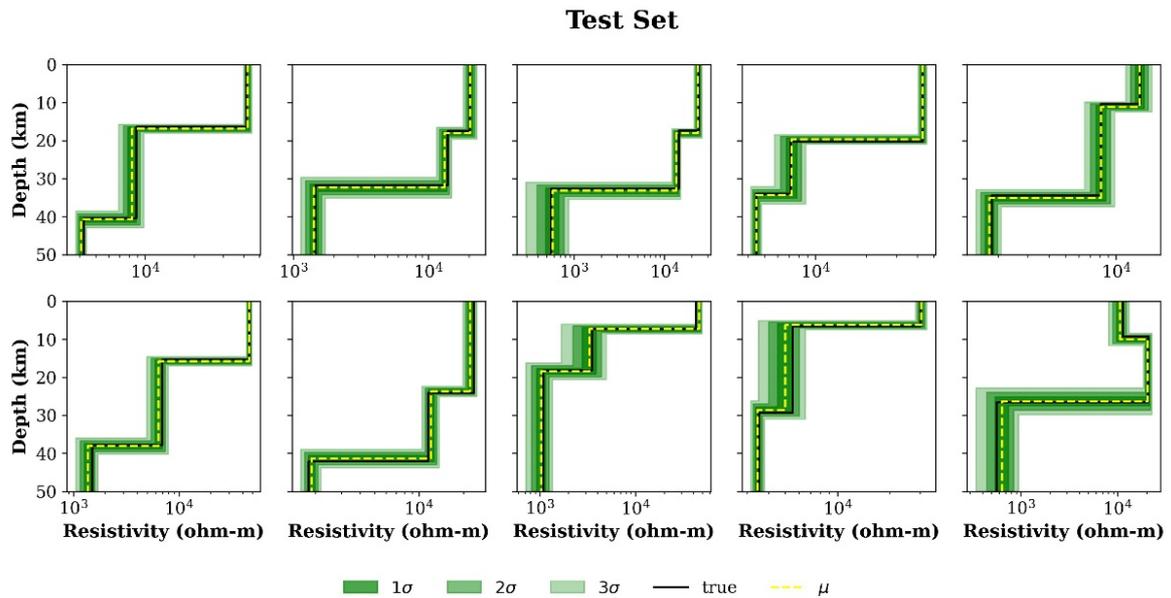

**Figure 3** Estimated resistivity profiles from the BCNN model (Figure 1) for selected test set examples. True profiles are shown in black, mean predictions in yellow, and green shaded regions represent uncertainty bounds corresponding to ±3 standard deviations around the mean.

## Results and Discussion

The MT dataset was normalized using z-score normalization, and the BCNN architecture (Figure 1) was trained for 2000 epochs using the Adam optimizer, with a learning rate of 0.01 and a batch size of 128. A summary of the model's performance is provided in Table 1, and results on a few randomly selected training and test examples are illustrated in Figures 2 and 3. The root mean squared error (RMSE) between the predicted resistivity profiles (computed as the mean of posterior samples) and the true profiles, remained low—indicating strong predictive accuracy of the BCNN model. Furthermore, in regions where there was a misalignment between the predicted and true profiles, the model exhibited increased uncertainty, with the true profiles still falling within the predicted uncertainty bounds. This behaviour is indicative of well-calibrated uncertainty estimates and reinforces the reliability of the model's outputs. To further assess robustness, the sae architecture was trained on the same dataset perturbed with 2.5% Gaussian noise. The model continued to perform well, showing such that the true profiles remained within the uncertainty bounds—suggesting that the BCNN is resilient to noise and potentially suitable for field data scenarios where noise is unavoidable. Motivated by these results, future work will explore extending this framework to joint inversion of multiple geophysical datasets. We also aim to investigate how incorporating additional modalities, such as training with physics, affects uncertainty estimates, and ultimately apply this approach to a field case study to validate its real-world applicability.

## Conclusions

This study demonstrates the potential of BNNs as a principled framework for magnetotelluric (MT) inversion with integrated uncertainty quantification. The proposed BCNN architecture exhibited strong predictive performance, with the predicted means closely matching the true resistivity profiles across both the training and test sets. Notably, the true profiles consistently fell within the predicted uncertainty bounds—even in regions where the predicted means deviated from the ground truth. These findings underscore the efficacy of BNNs not only in recovering geophysical parameters but also in capturing the uncertainty inherent in inverse problems addressed through supervised deep learning.